\documentclass[fleqn,usenatbib]{mnras}

\usepackage[T1]{fontenc}

\DeclareRobustCommand{\VAN}[3]{#2}
\let\VANthebibliography\thebibliography
\def\thebibliography{\DeclareRobustCommand{\VAN}[3]{##3}\VANthebibliography}

\usepackage{graphicx}	
\usepackage{amsmath}	
\usepackage{amssymb}	

\title[Origin of slow-drift shadow bursts in Jovian  DAM]{Origin of slow-drift shadow bursts in Jovian  decameter radio emission with quasi-harmonic structure}

\author[V. E. Shaposhnikov and V. V. Zaitsev]{
V.E. Shaposhnikov,$^{1,2}$\thanks{E-mail: sh130@ipfran.ru}
V.V. Zaitsev,$^{1,3}$
\\
$^{1}$Institute of Applied Physics of the Russian Academy of Sciences, Nizhny Novgorod,
Russia\\
$^{3}$High School of Economics, National Research University, Nizhny Novgorod Branch, Nizhny Novgorod, Russia\\
$^{2}$Pulkovo Observatory of Russian Academy of Sciences, Saint-Petersburg, Russia}

\pubyear{2023}

\begin{document}
\label{firstpage}
\pagerange{\pageref{firstpage}--\pageref{lastpage}}
\maketitle

\begin{abstract}

An explanation is proposed for the appearance of slowly drift shadow bursts in the dynamic spectrum of Jupiter against the background of decameter radio emission with a quasi-harmonic structure. Background radio emission is caused by hot ions with a loss cone type distribution function, which generate ion cyclotron waves due to the effect of double plasma resonance. A flow of hot ions with a distribution function of the Maxwell type is injected into the source region, fills the loss cone of generating ions and interrupts the generation of ion cyclotron waves due to the filling of the loss cone. The condition under which instability breaks down is obtained, and the optimal values of the parameters of the injected ions necessary for the occurrence of bursts in absorption are determined.

\end{abstract}




\section{Introduction}

Long-term observations of decameter radio emission from Jupiter have revealed a rich time-frequency structure of its dynamic spectra
 \citep[see, for example,][and the literature cited there]{Zarka(1998),Litvinenko(llrkrtvs)(2009),Litvinenko(lskzpdbrvm)(2016),Panchenko(prrbzlskmfs)(2018)}. In particular, so-called "bursts in absorption" (or ''shadow bursts'') were observed. These bursts are light bands drifting in frequency, crossing narrow-band continuous radiation \citep{Riihimaa(rcfggll)(1981),Koshida(koik)(2010),Litvinenko(lskzpdbrvm)(2016),Shaposhnikov(slzzk)(2021)}.
  \begin{figure}
	\includegraphics[width=\columnwidth]{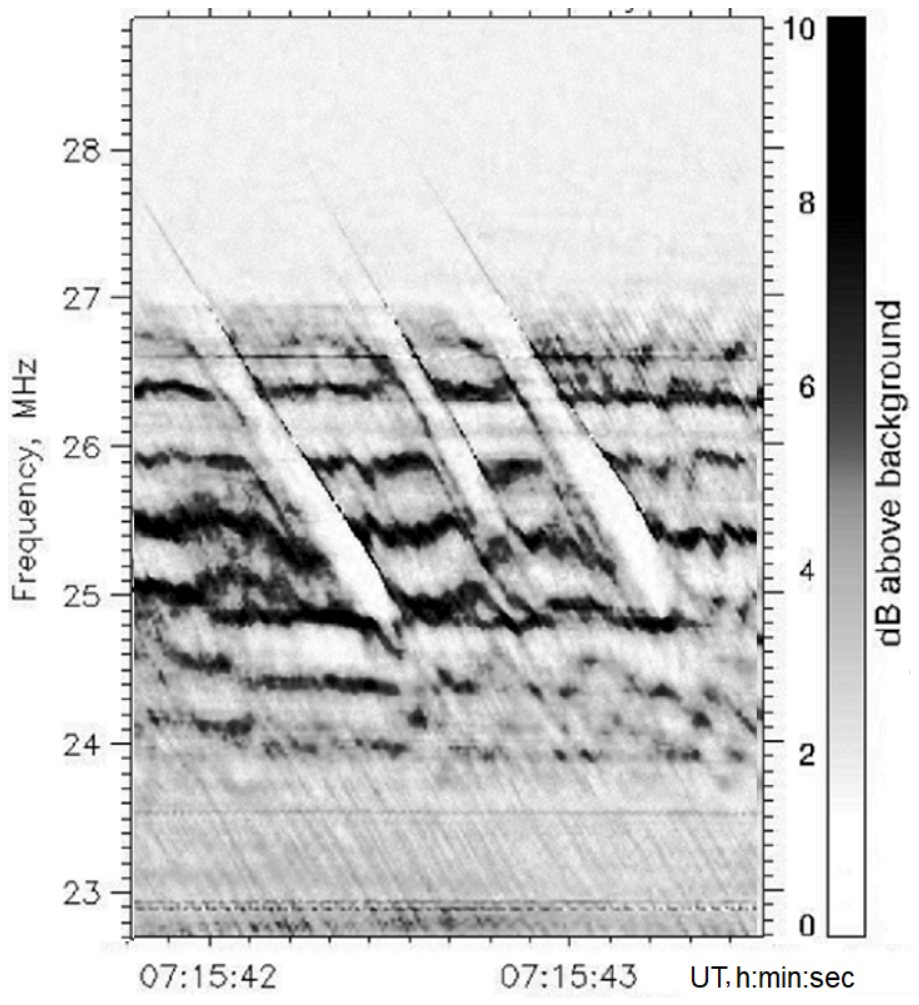}
    \caption{fragment of the dynamic spectrum of Jovian decameter radio emission from the Io-A source, in which bursts in absorption intersect continuous emission with a quasi-harmonic structure (this figure is adapted from Figure 1 in \citet{Shaposhnikov(slzzk)(2021)}).  }
    \label{fig1}
\end{figure}
Figure~\ref{fig1} shows a fragment of the dynamic spectrum of Jovian decameter radio emission from the Io-A source, in which bursts in absorption intersect continuous emission with a quasi-harmonic structure (this figure is adapted from Figure~\ref{fig1} in \citet{Shaposhnikov(slzzk)(2021)}). These bursts show a negative frequency drift with a drift rate of $\approx-4$~MHz/s. Note that similar bursts in absorption with a drift velocity of $\approx -5$~MHz/s were observed against the background of L-emission from the Io-A source at the Iitate Observatory (Tokyo) \citep{Koshida(koik)(2010)}. The authors called this phenomenon ''slow shadow bursts'', in contrast to the shadow bursts  reported in \citet{Riihimaa(rcfggll)(1981)}. Those bursts have a significantly higher drift speed, comparable to the speed of S-bursts. \citet{Gopalswamy(1986)} interprets the appearance of such shadow bursts as a result of the interaction of electron flows generating S- and L-bursts.

The mechanism for the formation of absorption bursts observed against the background of a burst of decameter radio emission with a quasi-harmonic structure is different. The fact is that, as shown by the analysis carried out in the work of \citet{Shaposhnikov(slzzk)(2021)}, electron beams are unable to generate decameter radio emission with such a frequency structure. The generation of such radiation occurs due to a plasma mechanism based on the effect of double plasma resonance\footnote{Here the effect of double plasma resonance is an increase in the growth rate by one to two orders of magnitude when the frequency of the lower hybrid resonance coincides with one of the ion cyclotron harmonics.}  on ion cyclotron harmonics, where fluxes of fast ions with a velocity distribution function of the ''loss cone'' type excite ion cyclotron waves with their subsequent conversion into electromagnetic radiation. The burst generation region here turns out to be elongated along the lines of force of the planetary magnetic field, and the generation of different quasi-harmonic radiation bands occurs in sources spaced apart in height, the position of which is determined by the condition that the frequency of the lower hybrid resonance coincides with one of the ion cyclotron harmonics.

\section{Slow shadow bursts formation model.}

The source of the burst with a quasi-harmonic structure and containing slow shadow bursts is a region of the planet's magnetosphere limited in height, filled with background equilibrium plasma, which determines the dispersion properties of excited ion cyclotron waves and converted electromagnetic waves. In this region there are quasi-stationary suprathermal nonequilibrium ions with a distribution function of the ''loss cone'' type, exciting ion cyclotron waves in the source areas where the condition of double plasma resonance with one of the ion cyclotron harmonics is satisfied. An upward flow of ions is injected into the burst generation region and fills the ''loss cone'' of emitting ions.  The ion flow fills the ''loss cone'' of emitting ions, as a result of which the instability breaks down, the generation of ion cyclotron waves stops, and a shadow burst appears on the dynamic spectrum. The injection velocity determines the drift rate of the shadow burst.

Let us consider a system of ions consisting of three components
\begin{equation}
f(v_{\parallel},v_{\perp})=\sum_{\alpha=0}^{\alpha=2}f_{\alpha}(v_{\parallel},v_{\perp}),
\label{1}
\end{equation}
where $v_{\parallel}$ and $v_{\perp}$ are the longitudinal and transverse components of the ion velocity relative to the direction of the planet's magnetic field, $f_{0}$ is the equilibrium component of the plasma
\begin{equation}
f_{0}(v_{\parallel},v_{\perp})=\frac{N_{0}}{(\sqrt{2\pi})^3v^3_{\rm T}}\exp{\left( -\frac{v_{\parallel}+v_{\perp}}{2v_{\rm T}^2}\right)},
\label{2}
\end{equation}
$v_{\rm T}$ is the thermal velocity of equilibrium ions, $N_{0}$ is their density, $f_{1}$ is the distribution function of ions generating ion cyclotron waves
\begin{equation}
f_{1}(v_{\parallel},v_{\perp})=\frac{N_{1}v_{\perp}^{2}}{2(\sqrt{2\pi})^3 a_{1}^3}\exp\left(-\frac{v_{\parallel}+v_{\perp}}{2a_{1}^2}\right).
\label{3}
\end{equation}
The density of generating ions $N_{1}$ is much less than the concentration of equilibrium ions, $N_1\ll N_{0}$, and the characteristic velocity $a_{1}$ is much greater than the thermal one, $a_{1} \gg v_{\rm T}$. Under these conditions, nonequilibrium ions do not affect the dispersion properties of waves, and the absorption of generated waves by equilibrium particles can be neglected. For simplicity, we take the distribution function of injected ions in the form Maxwellian function
\begin{equation}
f_{2}(v_{\parallel},v_{\perp})=\frac{N_{2}\left(x-v_{\parallel}t \right)}{2(\sqrt{2\pi})^3 a_{2}^3}\exp\left(-\frac{v_{\parallel}+v_{\perp}}{2a_{2}^2}\right).
\label{4}
\end{equation}
where $x$ is the coordinate along the direction of propagation of the flow (along the magnetic field), the dependence of the concentration $N_2$ on the variable $x-v_\parallel t$ describes the propagation of the ion flow along the magnetic field. We consider the concentration of injected ions to be small, $N_2 \ll N_0$ and $a_2 \gg v_T$.

In an equilibrium plasma $f_0(v_\parallel,v_\perp)$ with an admixture of nonequilibrium particles with distribution functions $f_1(v_\parallel,v_\perp)$ and $f_2(v_\parallel,v_\perp)$, the dispersion relation for ion cyclotron waves can be obtained from the equality
\begin{equation}
\varepsilon_{\parallel}^{(0)}+\varepsilon_{\parallel}^{(1)}+\varepsilon_{\parallel}^{(2)}=0,
\label{5}
\end{equation}
where $\varepsilon_{\parallel}^{(0)}$  and $\varepsilon_{\parallel}^{(1.2)}$ are longitudinal, with respect to the direction of the magnetic field, components of the plasma dielectric permittivity, which are due to equilibrium $(\varepsilon_{\parallel}^{(0)})$ and nonequilibrium $(\varepsilon_{\parallel}^{(0)})$ particles. Taking into account (\ref{5}), the growth rate of instability of ion cyclotron waves can be written as follows
\begin{equation}
\gamma  \approx - \frac{{\rm Im}\,\epsilon_{\parallel}^{(1)}+{\rm Im}\,\epsilon_{\parallel}^{(2)}}{\left[\displaystyle {\partial \over \partial\omega}\epsilon_{\parallel}^{(0)}\right]_{\epsilon_{\parallel}^{(0)}=0}}.
\label{6}
\end{equation}
The amplification (absorption) of the wave with time $t$, due to the presence of suprathermal ions in the source, depends on the sign of the growth rate $\gamma $ (the amplitude of the wave is proportional to $A \propto e^{-\gamma t}$), i.e. on the sign of the numerator ${\rm Im}\,\epsilon_{\parallel}^{(1)}+{\rm Im}\,\epsilon_{\parallel}^{(2)}$,
\begin{eqnarray}
{\rm Im}\,\epsilon_{\parallel}^{(1)}+{\rm Im}\,\epsilon_{\parallel}^{(2)}=-\frac{2\pi^2}{k^2}\sum_{\alpha=1}^{\alpha=2}\omega_{\alpha}^2\sum_{l=-\infty}^{l=+\infty}\int_{-\infty}^{+\infty}
dv_{\parallel}\int_{0}^{+\infty}dv_{\perp}  \nonumber \\
J^2_{l}\left(\frac{k_{\perp}v_{\perp}}{\omega_{\rm B}} \right)
\times \left(k{\parallel}v_{\perp}\frac{\partial f_{\alpha}}{\partial v_{\parallel}}+l\omega_{\rm B}\frac{\partial f_{\alpha}}{\partial v_{\perp}} \right)
\delta (\omega -l\omega_{\rm B}-k_{\parallel}v_{\perp}).
\label{7}
\end{eqnarray}
In (\ref{7}) $J_{l}\left(\frac{k_{\perp} v_{\perp}}{\omega_{\rm B}} \right)$ - Bessel function of order $l$, $\omega_{1,2}=\sqrt{4\pi e^2 N_{1,2}/M}$, $\omega_{\rm B}=eB/Mc$ - plasma frequency and gyrofrequency of ions, respectively, $e$ and $M$ charge and mass of the ion.
When assessing the conditions under which instability breaks down at the double plasma resonance and the shadow burst is observed in the dynamic spectrum, we can neglect for simplicity the movement of ions with the distribution function $f_{2}(v_{\parallel},v_{\perp})$ along the magnetic field, i.e. neglect the time dependence of the $N_2$ ion concentration. The latter can be done provided
\begin{equation}
\left| \frac{\partial f_{2}}{\partial v_{\parallel}} \right|\gg t\left|\frac{\partial f_{2}}{\partial \zeta} \right|
\label{8}
\end{equation}
where $ \zeta = x-v_{\parallel}t $.

According to \citet{Shaposhnikov(slzzk)(2021)}, a significant enhancement of ion cyclotron waves caused by nonequilibrium particles with the distribution function $f_{1}(v_{\parallel},v_{\perp})$ due to the effect of double plasma resonance occurs in the hybrid frequency band\footnote{The hybrid band is the frequency band between adjacent harmonics of the gyrofrequency $l \omega_{\rm B}$ and $(l+1)\omega_{\rm B}$, containing the frequency of the lower hybrid resonance.}. Outside this band, the growth rate of instability of these waves is one to two orders of magnitude smaller. In addition, within
the hybrid band the gain is not the same; for waves whose frequency is adjacent to the lower boundary of the band, the gain is significantly higher. Substituting into (\ref{7}) the distribution functions presented in (\ref{3}) and (\ref{4}), and taking into account the above, we obtain the following expression for the imaginary part of the dielectric constant ${\rm Im}\,\epsilon_{\parallel}^{(1)}+{\rm Im}\,\epsilon_{\parallel}^{(2)}$
\begin{eqnarray}
{\rm Im}\,\epsilon_{\parallel}^{(1)}+{\rm Im}\,\epsilon_{\parallel}^{(2)} \approx \sqrt{\frac{\pi}{2}}\frac{\omega^2_{1}\omega_{\rm B}}{k_{
\parallel}k^2a_{1}^2}l \nonumber \\
\left[\exp \left(-Z^2_{l,1}\right) \xi_1 \varphi_{l}'(\xi_1)+\frac{N_2}{N_1}\left(\frac{a_1}{a_2} \right)^3\exp \left(-Z^2_{l,2} \right)\varphi_l(\xi_2)  \right]
\label{9}
\end{eqnarray}
where $\varphi_l(\xi)=e^{-\xi}I_{l}(\xi)$; $I_{l}(\xi)$ - modified Bessel function of order {l}; $\xi_{1,2}=k_{\perp}^2a_{1,2}^2/\omega_{\rm B}^2$;  $Z_{l,\alpha} =(\omega-l\omega_{\rm B})/\sqrt{2}k_{\parallel}a_{\alpha}$.

\begin{figure}
	\includegraphics[width=\columnwidth]{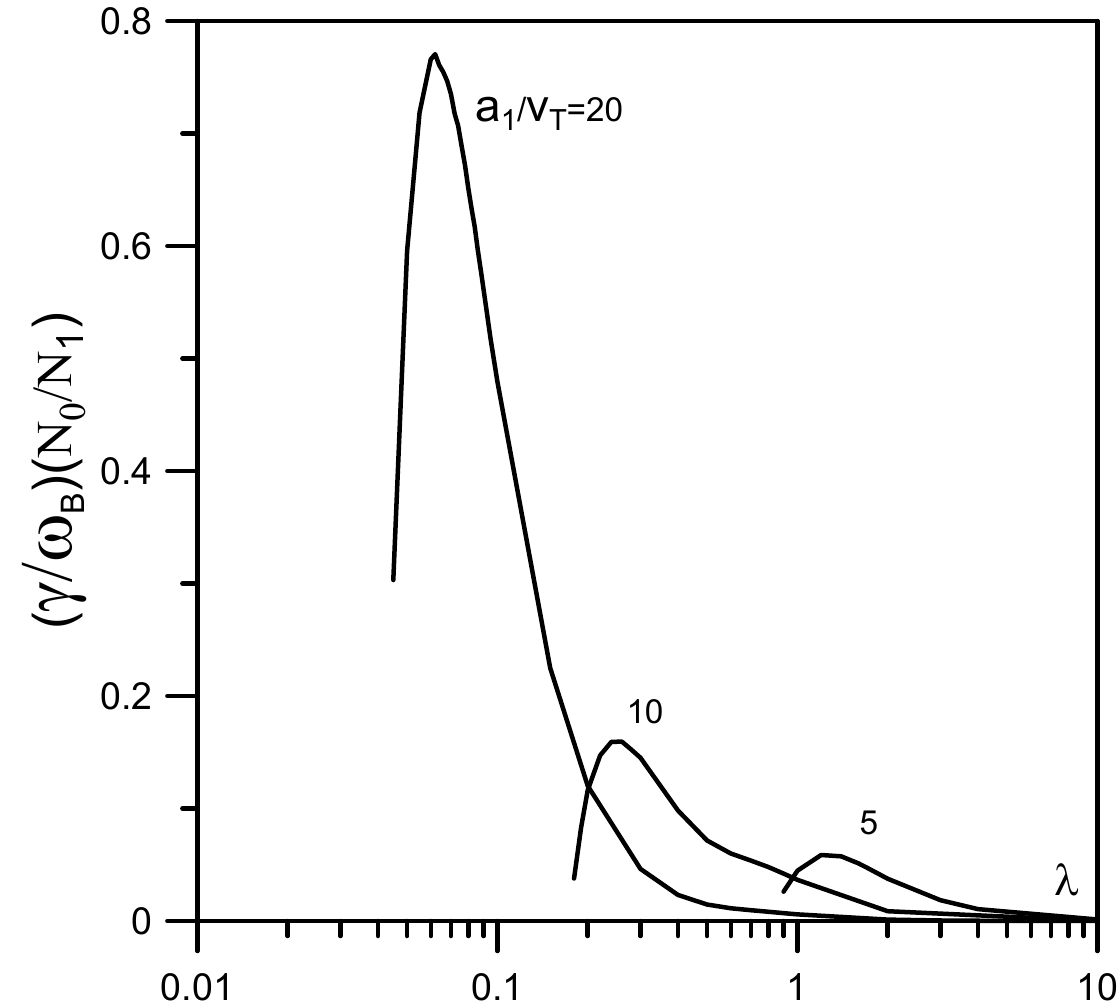}
    \caption{Instability of growth rate of ion cyclotron waves in the hybrid band $4\leq l\leq 5$  (this figure is adapted from Figure 4 in \citet{Shaposhnikov(szl)(2018)}).  }
    \label{fig2}
\end{figure}

In the absence of ion flow filling the loss cone, $N_2=0$, ion cyclotron waves are excited by nonequilibrium ions with a distribution function $f_1(v_{\parallel},v_{\perp})$ due to the effect of double plasma resonance. According to \citet{Shaposhnikov(slzzk)(2021)}, at a fixed frequency the maximum increment is achieved at $Z_{l,\alpha} =(\omega-l\omega_{\rm B})/\sqrt{2}k_{\parallel}a_{\alpha}\approx 1/\sqrt{2}$, i.e. in the direction $k_{\parallel} \approx (\omega-l\omega_{\rm B})/a_1$. Injection of ions into the radiation generation region leads to disruption of instability in this direction and the appearance of shadow burst, provided that the density of injected ions  $N_2$  exceeds a certain value
\begin{equation}
\frac{N_2}{N_1}>\frac{1}{\sqrt{e}}\left(\frac{a_2}{a_1} \right)^3\exp \left(\frac{a_1^2}{2a_2^2} \right)\lambda_{\max}\frac{a_1^2}{v_{\rm T}^2}
\frac{\phi_{l}'\left(\lambda_{\max}\frac{a_1^2}{v_{\rm T}^2} \right)}{\phi_{l}\left(\lambda_{\max}\frac{a_1^2}{v_{\rm T}^2 \left(\frac{a_2}{a_1}\right)^2} \right)},
\label{10}
\end{equation}
where $\lambda_{\max}=k^2_{\perp,\max}v_{\rm T}^2/\omega_{\rm B}^2 $ is the parameter at which the cyclotron instability growth rate reaches its maximum value (see Fig.~\ref{fig2}).

In the framework of the model of a radio source with a quasi-harmonic structure based on the double resonance effect, the number of the hybrid band is a free parameter. Figure~\ref{fig2} (adapted from Figure 4 in \citet{Shaposhnikov(szl)(2018)} shows an example of the dependence of the instability increment of cyclotron waves in the hybrid band $4 \leq l \leq 5$ for different values of the ratio of the characteristic velocity of emitting ions to the thermal velocity of ions of equilibrium plasma from the wave vector of excited waves. The increment reaches its maximum value at $\lambda_{\max}\simeq 0.06; 0.2; 1.2$ when the ratio of the average speed of emitting electrons to the thermal speed of the background plasma is  $a_1/v_{\rm T}=20; 10; 5$ , respectively.
\begin{figure}
	\includegraphics[width=\columnwidth]{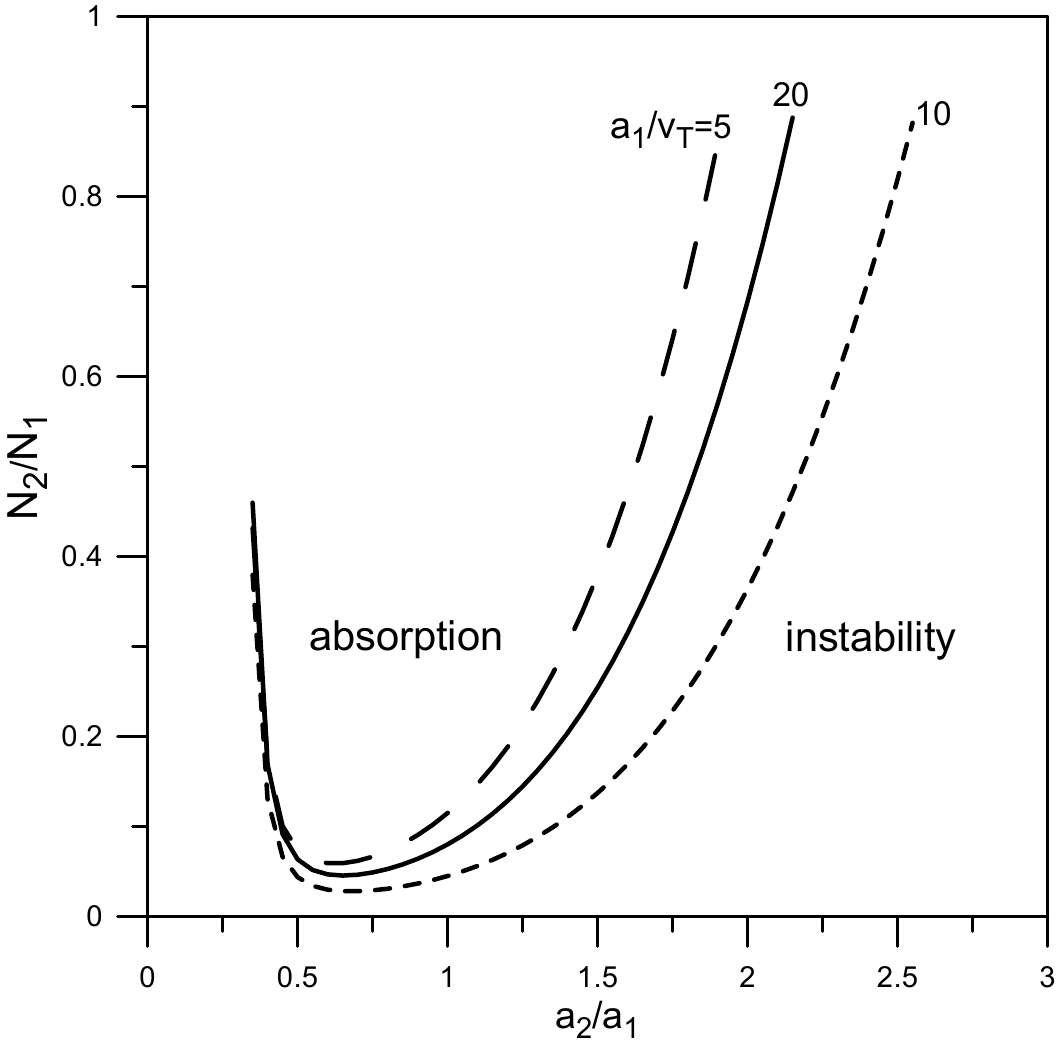}
    \caption{Dependence of the limiting density of injected ions depending on the ratio $a_2/a_1$.  }
    \label{fig3}
\end{figure}
For the same conditions Figure~\ref{fig3}  shows the limiting density of injected ions depending on the ratio $a_2/a_1$, above which the instability breaks down and shadow burst occurs.

\section*{Acknowledgements}
The study was carried out with financial support from the Russian Science Foundation (grant 20-12-00268-P).


\begin{thebibliography}{}
\makeatletter
\relax
\def\mn@urlcharsother{\let\do\@makeother \do\$\do\&\do\#\do\^\do\_\do\%\do\~}
\def\mn@doi{\begingroup\mn@urlcharsother \@ifnextchar [ {\mn@doi@}
  {\mn@doi@[]}}
\def\mn@doi@[#1]#2{\def\@tempa{#1}\ifx\@tempa\@empty \href
  {http://dx.doi.org/#2} {doi:#2}\else \href {http://dx.doi.org/#2} {#1}\fi
  \endgroup}
\def\mn@eprint#1#2{\mn@eprint@#1:#2::\@nil}
\def\mn@eprint@arXiv#1{\href {http://arxiv.org/abs/#1} {{\tt arXiv:#1}}}
\def\mn@eprint@dblp#1{\href {http://dblp.uni-trier.de/rec/bibtex/#1.xml}
  {dblp:#1}}
\def\mn@eprint@#1:#2:#3:#4\@nil{\def\@tempa {#1}\def\@tempb {#2}\def\@tempc
  {#3}\ifx \@tempc \@empty \let \@tempc \@tempb \let \@tempb \@tempa \fi \ifx
  \@tempb \@empty \def\@tempb {arXiv}\fi \@ifundefined
  {mn@eprint@\@tempb}{\@tempb:\@tempc}{\expandafter \expandafter \csname
  mn@eprint@\@tempb\endcsname \expandafter{\@tempc}}}

\bibitem[\protect\citeauthoryear{Gopalswamy}{Gopalswamy}{1986}]{Gopalswamy(198%
6)}
Gopalswamy N.,  1986, \mn@doi [Earth, Moon, and Planets] {10.1007/BF00117270},
  35, 93

\bibitem[\protect\citeauthoryear{Koshida, Ono, Iizima  \& Kumamoto}{Koshida
  et~al.}{2010}]{Koshida(koik)(2010)}
Koshida T.,  Ono T.,  Iizima M.,   Kumamoto A.,  2010, \mn@doi [\jgr]
  {10.1029/2009JA014608}, 115, A01202

\bibitem[\protect\citeauthoryear{Litvinenko, Rucker, Vinogradov, Leiner  \&
  Shaposhnikov}{Litvinenko et~al.}{2009}]{Litvinenko(llrkrtvs)(2009)}
Litvinenko G.~V.,  Rucker H.~O.,  Vinogradov V.~V.,  Leiner M.,   Shaposhnikov
  V.~E.,  2009, \mn@doi [Astron. Astrophys.] {10.1051/0004-6361:200809676},
  493, 651

\bibitem[\protect\citeauthoryear{Litvinenko et~al.,}{Litvinenko
  et~al.}{2016}]{Litvinenko(lskzpdbrvm)(2016)}
Litvinenko G.~V.,  et~al., 2016, \mn@doi [Icarus]
  {10.1016/j.icarus.2016.02.039}, 272, 80

\bibitem[\protect\citeauthoryear{Panchenko et~al.,}{Panchenko
  et~al.}{2018}]{Panchenko(prrbzlskmfs)(2018)}
Panchenko M.,  et~al., 2018, \mn@doi [Astron. Astrophys.]
  {10.1051/0004-6361/201731369}, 610, A69

\bibitem[\protect\citeauthoryear{Riihimaa, Carr, Flagg, Greenman, Gombola, Lebo
   \& Levy}{Riihimaa et~al.}{1981}]{Riihimaa(rcfggll)(1981)}
Riihimaa J.~J.,  Carr T.~D.,  Flagg R.~S.,  Greenman W.~B.,  Gombola P.~P.,
  Lebo G.~R.,   Levy J.~A.,  1981, \mn@doi [Icarus.]
  {10.1016/0019-1035(81)90111-1}, 48, 298

\bibitem[\protect\citeauthoryear{Shaposhnikov, Zaitsev  \&
  Litvinenko}{Shaposhnikov et~al.}{2018}]{Shaposhnikov(szl)(2018)}
Shaposhnikov V.~E.,  Zaitsev V.~V.,   Litvinenko G.~V.,  2018, \mn@doi [J.
  Geophys. Res.] {10.1029/2018JA026064}, 123, 9395

\bibitem[\protect\citeauthoryear{Shaposhnikov, Litvinenko, Zaitsev, Zakharenko
  \& Konovalenko}{Shaposhnikov et~al.}{2021}]{Shaposhnikov(slzzk)(2021)}
Shaposhnikov V.~E.,  Litvinenko G.~V.,  Zaitsev V.~V.,  Zakharenko V.~V.,
  Konovalenko A.~A.,  2021, \mn@doi [\aap] {10.1051/0004-6361/202039304}, 645,
  A31

\bibitem[\protect\citeauthoryear{Zarka}{Zarka}{1998}]{Zarka(1998)}
Zarka P.,  1998, \jgr, 103, 20159

\makeatother
\end{thebibliography}

\label{lastpage}
\end{document}